\documentclass[doublecol,figures]{epl2}
\usepackage{amsmath}
\usepackage{amssymb}

\title{Operator Quantum Geometric Tensor and Quantum Phase Transitions}

\author{Xiao-Ming Lu \and Xiaoguang Wang \thanks{E-mail: \email{xgwang@zimp.zju.edu.cn}}}
\shortauthor{X.-M Lu \etal}

\institute{Zhejiang Institute of Modern Physics, Department of Physics, Zhejiang
University, Hangzhou 310027, P.R. China.}


\abstract{
We extend the quantum geometric tensor from the state space to the
operator level, and investigate its properties like the additivity
for factorizable models and the splitting of two kinds contributions
for the case of stationary reference states. This operator-quantum-geometric
tensor (OQGT) is shown to reflect the sensitivity of unitary operations
against perturbations of multi parameters. General results for the
cases of time evolutions with given stationary reference states are
obtained. By this approach, we get exact results for the rotated XY models,
and show relations between the OQGT and quantum criticality.
}

\begin{document}
\maketitle

\section{Introduction}

The geometric structure on the manifold of quantum states (MQS) is
an interesting subject of quantum physics, which was studied from
different perspectives, e.g., the geometrization of quantum mechanics
\cite{Kibble1979}, the geometric phase \cite{Berry1984,Simon1983,Aharonov1987,Shapere1989},
the geometry of quantum evolution \cite{Anandan1990}, the quantum
theory of gravity \cite{Ashtekar1999,Minic2003,Carroll2005}, quantum
phase transitions \cite{Zanardi2007,Venuti2007}, etc. Generally speaking,
on the MQS, there is a natural Riemann structure---the quantum-geometric
tensor (QGT), induced by the inner product in the Hilbert space $\mathcal{H}$
\cite{Provost1980,Shapere1989}. The real part of the QGT provides
the metric tensor, through which we can measure the distance between
two states on the MQS. The imaginary part of QGT gives a curvature
2-form, corresponding to a natural symplectic structure on the projective
Hilbert space $\mathcal{P}(\mathcal{H})$. With this curvature 2-form,
the geometric phase can be interpreted as the area enclosed by the
closed curve in $\mathcal{P}(\mathcal{H})$ \cite{Anandan1988}.

The corresponding metric given by the QGT is a kind of Fubini-Study
metric related to the fidelity of two quantum states, which is one
of the most popular physical quantities in the quantum-information
processing. Recently, it is proposed that the fidelity can characterize
quantum phase transitions \cite{Zanardi2006}. This approach has been
achieved a lot of progress \cite{Zanardi2006,Zhou2008a,Zanardi2007,%
Zanardi2007a,Zanardi2007b,Zhou2008,Venuti2007,Cozzini2007,You2007,
Buonsante2007,Chen2007,Chen2008,Yang2007,Yang2008,Gu2008,Ma2008,Kwok2008,Gu2008a,Quan2009}.
More recently, the fidelity approach was extended to the operator
level---the operator fidelity (OF) \cite{Wang2009,Lu2008}. The underlying
idea behind the OF is mapping the unitary operators into the quantum
states. It is proposed that the susceptibility (metric) of the OF
can be used to study the stability of quantum evolution with respect
to perturbations, indicate the quantum critical points, quantify the
environment induced decoherence and investigate quantum chaos \cite{Wang2009,Lu2008,Sun2009,Giorda2010,TobiasJacobson2010}.

Since unitary operators can be mapped into quantum states, on
the manifold of the unitary operators , we can define an operator-quantum-geometric
tensor (OQGT), whose real and imaginary parts can give a metric tensor
and a curvature 2-form respectively. Like the QGT is naturally induced
by the inner product in the Hilbert space $\mathcal{H},$ the OQGT
is naturally induced by the inner product of two operators, which
is a generalization of the Hilbert-Schmidt product, with a reference
state $\rho$ \cite{Lu2008}. Like the susceptibility of OF, The OQGT has
the additivity for the factorizable model and the splitting of two kinds of contributions
for the special cases that the reference state $\rho$ is commutable
with the operators. We will show this OQGT approach is useful to the sensitivity of unitary operations against multi-parameter perturbations.

\section{Riemannian structure on the MQS}

The term MQS is pointed to the submanifold of the projective Hilbert
space, which is defined as the sets of rays of the Hilbert space $\mathcal{H}$.
The inner product structure of the Hilbert space $\mathcal{H}$ naturally
induce a Riemannian structure of the projective Hilbert space $\mathcal{P}(\mathcal{H})$
\cite{Provost1980}. We denote the ray of $\mathcal{H}$ by $|\tilde{\psi}\rangle$.
The Fubini-Study metric indicating the distance between $|\tilde{\psi}\rangle$
and $|\tilde{\phi}\rangle$ is defined as follows\begin{equation}
\gamma(\psi,\phi)=\arccos F,\end{equation}
 where $F=|\langle\psi|\phi\rangle|$ is the fidelity defined through
the inner product in $\mathcal{H}$. The infinitesimal form of this
metric, which denote the distance between two slightly different states
$|\tilde{\psi}(\lambda)\rangle$ and $|\tilde{\psi}(\lambda+d\lambda)\rangle$,
can be written as\begin{equation}
ds^{2}\simeq4(1-F^{2})=Q_{\mu\nu}d\lambda^{\mu}d\lambda^{\nu},\label{eq:distance}\end{equation}
 where \begin{equation}
Q_{\mu\nu}\equiv\left\langle \partial_{\mu}\psi\left|(1-|\psi\rangle\langle\psi|)\right|\partial_{\nu}\psi\right\rangle \end{equation}
 is the so-called quantum geometric tensor on $\mathcal{P}(\mathcal{H})$
\cite{Provost1980,Shapere1989}. Here $\partial_{\mu}\equiv\partial/\partial\lambda^{\mu}$
with $\lambda^{\mu}$-s the parameters characterizing the elements
in $\mathcal{P}(\mathcal{H})$. Hereafter, we use the Einstein summation
convention. $Q_{\mu\nu}$ is a Hermitian matrix, whose real part is
symmetric while the imaginary part is antisymmetric. Hence, only the
real part of quantum geometric tensor has effects on Eq.~(\ref{eq:distance})
due to the summation , i.e.,$ds^{2}=g_{\mu\nu}d\lambda^{\mu}d\lambda^{\nu}$,
where $g_{\mu\nu}=\mathrm{Re}\, Q_{\mu\nu}$. Meanwhile, the imaginary
part gives the curvature 2-form\begin{eqnarray}
\sigma & = & \left(\mathrm{Im}\, Q_{\mu\nu}\right)d\lambda^{\mu}\wedge d\lambda^{\nu},\end{eqnarray}
 where $\wedge$ is the exterior (wedge) product. The integration
of $\sigma$ over an area $S$ gives the geometric phase $\gamma_{g}=-\iint_{S}\sigma$
of the cycle evolution along the boundary of the area $S$ \cite{Berry1984,Aharonov1987}.
Further, $\sigma$ can be written as the exterior differential of
a 1-form---Berry's connection $\beta=-i\langle\psi|d\psi\rangle$,
i.e., $\sigma=d\beta=-i\langle d\psi|\wedge|d\psi\rangle$.

\section{Operator quantum geometric tensor}

In the space of operators, one can define the following Hermitian
inner product
\begin{equation}
    \langle X,Y\rangle_{\rho}:=\mathrm{Tr}(X^{\dagger}Y\rho),\label{eq:operator_inner_prod}
\end{equation}
where $\rho$ is a density operator taken as the reference state.
Especially, we concern the unitary operators hereafter, which satisfies
the normalized condition $\langle U,U\rangle_{\rho}=1$. Actually,
this inner product structure can be interpreted as the conventional
one for the two pure states $|X\rangle_{\rho}$ and $|Y\rangle_{\rho}$
in the extended Hilbert space \cite{Lu2008}, through a map from the
operator $X$ to the pure state $|X\rangle_{\rho}$ given as follows
\begin{equation}
    X\mapsto|X\rangle_{\rho}\equiv X\otimes\mathbf{1}|\Psi(\rho)\rangle,
\end{equation}
where $\mathbf{1}$ is the identity operator in the ancillary Hilbert
space $\mathcal{H}^{\mathrm{anc}}$ and $|\Psi(\rho)\rangle$ is a
pure state satisfying $\mathrm{Tr_{anc}\left(|\Psi(\rho)\rangle\langle\Psi(\rho)|\right)=\rho}$,
i.e., it is a purification of $\rho$ in the extended Hilbert space.
Obviously, the concrete form of purification does not influence the
inner product (\ref{eq:operator_inner_prod}). For brevity, we omit
the subscript label $\rho$ of $|X\rangle_{\rho}$ hereafter as long
as not confusing.

Follow this spirit, the fidelity of two unitary operators is defined
by
\begin{equation}
    F(U_{1},U_{2})=|\mathrm{Tr}(U_{1}^{\dagger}U_{2}\rho)|,
\end{equation}
which is equivalent to the fidelity $|\langle U_{1}|U_{2}\rangle|$
of the two pure states $|U_{1}\rangle$ and $|U_{2}\rangle$. If $U_{1}$
and $U_{2}$ are slightly different and $\rho$ is pure, $F(U_{1},U_{2})$
is just the square root of the Loschmidt echo \cite{Gorin2006}, which
is used to describe the hypersensitivity of the time evolution to
perturbations. It is remarkable that the minimization of $F(U_{1},U_{2})$
with respect to a optimal $\rho$ is used to characterize the statistical
distinguishability of two unitary operations $U_{1}$ and $U_{2}$
\cite{Ac'in2001}.

Like the Riemannian structure on the MQS \cite{Provost1980}, we can
naturally induce the Riemann structure on the manifold of the unitary
operators along with the given reference states. The main geometric
object characterizes this structure is the quantum geometric tensor
of the unitary operators
\begin{align}
    Q_{\mu\nu} & \equiv\left\langle \partial_{\mu}U|\partial_{\nu}U\right\rangle -\left\langle \partial_{\mu}U|U\right\rangle \left\langle U|\partial_{\nu}U\right\rangle \nonumber \\
    & =\left\langle A_{\mu}A_{\nu}\right\rangle _{\rho}-\left\langle A_{\mu}\right\rangle _{\rho}\left\langle A_{\nu}\right\rangle _{\rho},\label{eq:OQGT}
\end{align}
where $A_{\mu}\equiv iU^{\dagger}\partial_{\mu}U$ is the Hermitian
generator of displacements in $\lambda^{\mu}$ and $\left\langle O\right\rangle _{\rho}\equiv\mathrm{Tr}\left[\rho O\right]$
denotes the expected value of the observable $O$ on the state $\rho$.
Hereafter, we call $Q_{\mu\nu}$ operator-quantum-geometric tensor
(OQGT). The real part $g_{u\nu}=\mathrm{Re}\, Q_{\mu\nu}$ is the
Fubini-Study metric tensor of the unitary operator space with the
reference state $\rho$, i.e., $ds^{2}=g_{u\nu}d\lambda^{\mu}d\lambda^{\nu}$,
which measure the statistical distance \cite{Wootters1981,Braunstein1994}
between the two states $U(\lambda)\otimes\mathbf{1}|\Psi(\rho)\rangle$
and $U(\lambda+d\lambda)\otimes\mathbf{1}|\Psi(\rho)\rangle$ in the
extended Hilbert space. Meanwhile, due to the anti-commutative law
of the exterior product, the imaginary part $\mathrm{Im}Q_{\mu\nu}=\mathrm{Im}\langle A_{\mu}A_{\nu}\rangle$
gives the curvature 2-form as follows
\begin{equation}
\sigma\equiv-i\langle A\wedge A\rangle_{\rho},\label{eq:curvature_form}\end{equation}
where $A\equiv A_{\mu}d\lambda^{\mu}=iU^{\dagger}dU$ is a differential
1-form. If the reference states $\rho$ are independent on $\lambda^{\mu}$-s,
$\sigma$ can be written as $\sigma=d\beta$, where $\beta\equiv-\langle A\rangle_{\rho}$
is the analog of Berry's connection. Because $\beta$ can be written
as $\beta=-i\langle\Psi(\rho)|(U^{\dagger}\otimes\mathbf{I})(dU\otimes\mathbf{I})|\Psi(\rho)\rangle$
with $\Psi(\rho)$ the purification of $\rho$, it is, in fact, the
Berry's connection in the extended Hilbert space. For a closed trajectory
$\partial S=\{\lambda(t):t\in[0,T)\}$ with $\lambda(0)=\lambda(T)$
in the parameter manifold, the geometric phase is given by $\gamma_{\mathrm{g}}=-\oint_{\partial S}\beta=-\iint_{S}\sigma$,
where $S$ is the area subtended by the closed trajectory $\partial S$.

Next, we investigate the properties of the OQGT for the cases of factorizable
models and of stationary reference states. These will be useful to
the calculations in the XY model presented later.

\subsection{Additivity for factorizable models}

We analyze the case in which the unitary operator $U(\lambda)$ and
the reference state $\rho(\lambda)$ have the same factorization structure
as follows\begin{equation}
U(\lambda)=\bigotimes_{k=0}^{M}U_{k}(\lambda),\quad\rho(\lambda)=\bigotimes_{k=0}^{M}\rho_{k}(\lambda).\end{equation}
This assumption is sensible when $U(\lambda)$ and $\rho(\lambda)$
are both generated by the same Hamiltonian. For instance, $U=\exp\left(-itH(\lambda)\right)$
is the time evolution generated by $H(\lambda)$ and $\rho(\lambda)$
is the Gibbs thermal state which is a mixture of the eigenstates of
$H(\lambda)$. After derivative with respect to $\lambda^{\mu}$ we
obtain
\begin{equation}
    \partial_{\mu}U(\lambda) = \sum_{l=0}^{M}[\bigotimes_{k=0}^{l-1}U_{k}(\lambda)]\otimes\partial_{\mu}U_{l}(\lambda)\otimes[\bigotimes_{k=l+1}^{M}U_{k}(\lambda)].
\end{equation}
Because $\text{Tr}(A\otimes B)=\text{Tr}(A)\text{Tr}(B)$ and $\text{Tr}(\rho_{l})=1$,
we get
\begin{align*}
    \langle\partial_{\mu}U|\partial_{\nu}U\rangle= & \sum_{l}\langle\partial_{\mu}U_{l}|\partial_{\nu}U_{l}\rangle_{l}+\sum_{l\neq l^{\prime}}M_{ll^{\prime}}^{\mu\nu},\\
    \langle\partial_{\mu}U|U\rangle\langle U|\partial_{\nu}U\rangle= & \sum_{l}\langle\partial_{\mu}U_{l}|U_{l}\rangle_{l}\langle U_{l}|\partial_{\nu}U_{l}\rangle_{l}+\sum_{l\neq l^{\prime}}M_{ll^{\prime}}^{\mu\nu},
\end{align*}
where $M_{ll^{\prime}}^{\mu\nu}=\text{Tr}_{l}[\rho_{l}(\partial_{\mu}U_{l})^{\dagger}U_{l}]\text{Tr}_{l^{\prime}}[\rho_{l^{\prime}}U_{l^{\prime}}^{\dagger}(\partial_{\nu}U_{l^{\prime}})]$,
and $\langle A|B\rangle_{l}=\mathrm{Tr}_{l}[\rho_{l}A^{\dagger}B]$,
the subscript $l$ means the $l$-th subspace. In such cases, we have
\begin{equation}
    Q_{\mu\nu}  =  \sum_{l}Q_{\mu\nu}^{l},
\end{equation}
with $Q_{\mu\nu}^{l}=\langle\partial_{\mu}U_{l}|\partial_{\nu}U_{l}\rangle_{l}-\langle\partial_{\mu}U_{l}|U_{l}\rangle_{l}\langle U_{l}|\partial_{\nu}U_{l}\rangle_{l}$.
Hence, for the model whose unitary operators and the reference state
$\rho$ have the same factorization structure, the OQGT is additive.

\subsection{Splitting for stationary reference states}

For matrix which can be diagonalized, the variance with respect to
the parameters can be decomposed into two parts, the variance of the
eigenvalues and of eigenvectors. A unitary operator $U(\lambda)$
can always be written in the form $U(\lambda)=S^{\dagger}(\lambda)U_{\mathrm{d}}(\lambda)S(\lambda)$,
where $U_{\mathrm{d}}(\lambda)$ is a diagonal unitary matrix and
$S(\lambda)$ is unitary. So the derivative of $U$ can be splitted
into two terms as follows
\begin{equation}
    \partial_{\mu}U  = D_{\mu}^{(1)}+D_{\mu}^{(2)}
\end{equation}
with $D_{\mu}^{(1)}\equiv S^{\dagger}\left(\partial_{\mu}U_{\mathrm{d}}\right)S$
and $D_{\mu}^{(2)}\equiv-i[U,\mathfrak{a}_{\mu}]$, where $\mathfrak{a}_{\mu}\equiv iS^{\dagger}\partial_{\mu}S$
is a Hermitian matrix. Note that we have $[D_{\mu}^{(1)},U]=0$ and
$[D_{\mu}^{(1)\dagger},U]=0$, since $D_{\mu}^{(1)}$ and $U$ can
be simultaneously diagonalized by the unitary operator $S(\lambda)$.
Correspondingly, we have $A_{\mu}=A_{\mu}^{(1)}+A_{\mu}^{(2)}$ with
$A_{\mu}^{(i)}=iU^{\dagger}D_{\mu}^{(i)}$ for $i=1,2$.

For cases in which the reference states $\rho(\lambda)$ is stationary
under the unitary operation $U(\lambda)$, i.e., $[\rho,U]=0$, we
chose $S(\lambda)$ as the unitary matrix diagonalizing simultaneously
$U(\lambda)$ and $\rho(\lambda)$. Then we have
\begin{align}
    \langle A_{\mu}^{(2)}\rangle_{\rho} & =\mathrm{Tr}\left\{ U^{\dagger}[U,\mathfrak{a}_{\mu}]\rho\right\} =0,\nonumber \\
    \langle A_{\mu}^{(1)\dagger}A_{\nu}^{(2)}\rangle_{\rho} & =-i\mathrm{Tr\left\{ D_{\mu}^{(1)\dagger}\left[U,\mathfrak{a_{\nu}}\right]\rho\right\} }=0.\label{eq:terms}
\end{align}
The second equality of second line of the above equations is due to
the commutation relation $[D_{\mu}^{(1)\dagger},U]=0$ and $[\rho,U]=0$.
Then substituting $A_{\mu}=A_{\mu}^{(1)}+A_{\mu}^{(2)}$ into Eq.~(\ref{eq:OQGT})
and combining Eq.~(\ref{eq:terms}), we obtain the splitted form
of the OQGT as follows
\begin{eqnarray}
    Q_{\mu\nu} & = & Q_{\mu\nu}^{(1)}+Q_{\mu\nu}^{(2)},\nonumber \\
    Q_{\mu\nu}^{(1)} & = & \langle A_{\mu}^{(1)}A_{\nu}^{(1)}\rangle_{\rho}-\langle A_{\mu}^{(1)}\rangle_{\rho}\langle A_{\nu}^{(1)}\rangle_{\rho},\label{eq:Q_sep}\\
    Q_{\mu\nu}^{(2)} & = & \langle A_{\mu}^{(2)}A_{\nu}^{(2)}\rangle_{\rho}.\nonumber
\end{eqnarray}
Note that $[A_{\mu}^{(1)},A_{\nu}^{(1)}]=0$ and the Hermitian of
$A_{\mu}^{(1)}$, so $Q_{\mu\nu}^{(1)}$ is real. Then, the curvature
2-form can be expressed as \begin{equation}
\sigma=-i\langle A^{(2)}\wedge A^{(2)}\rangle_{\rho}\end{equation}
 with $A^{(i)}\equiv A_{\mu}^{(i)}d\lambda^{\mu}$.

A motivation of such separation may be seen from time dependence of
$Q^{(1)}$ and $Q^{(2)}$ if we consider the time evolution operator
$U(\lambda)=\exp\left[-itH(\lambda)\right]$ \cite{Lu2008}. $\hbar=1$
is assumed hereafter. We assume the Hamiltonian $H(\lambda)$ and
the reference state $\rho(\lambda)$ can be simultaneously diagonalized
by unitary matrix $S(\lambda)$, i.e., $H_{\mathrm{d}}(\lambda)=S(\lambda)H(\lambda)S^{\dagger}(\lambda)$
and $\rho_{\mathbf{d}}=S(\lambda)\rho(\lambda)S^{\dagger}(\lambda)$
are both of diagonal form. Correspondingly, we have $U(\lambda)=S^{\dagger}(\lambda)U_{\mathrm{d}}(\lambda)S(\lambda)$
with $U_{\mathrm{d}}(\lambda)=\exp\left[-itH_{\mathrm{d}}(\lambda)\right]$.
Substituting the time evolution form of $U(\lambda)$ into Eq.~(\ref{eq:Q_sep}),
we obtain
\begin{eqnarray}
    Q_{\mu\nu}^{(1)} & = & \alpha_{\mu\nu}t^{2},\nonumber \\
    Q_{\mu\nu}^{(2)} & = & \sum_{ij}\beta_{\mu\nu}^{ij}\left\{ 1-\cos\left[\left(E_{i}-E_{j}\right)t\right]\right\} ,\label{eq:OQGT_time_evolution}
\end{eqnarray}
where $E_{i}$ is the eigenvalue of $H$. $\alpha_{\mu\nu}$, $\beta_{\mu\nu}^{ij}$
are both time-independent and defined by
\begin{eqnarray}
    \alpha_{\mu\nu} & \equiv & \mathrm{Tr}\left[\left(\partial_{\mu}H_{\mathrm{d}}\right)\left(\partial_{\nu}H_{\mathrm{d}}\right)\rho_{\mathrm{d}}\right]\nonumber \\
     && \mathrm{-\mathrm{Tr}\left[\left(\partial_{\mu}H_{\mathrm{d}}\right)\rho_{\mathrm{d}}\right]\mathrm{Tr}\left[\left(\partial_{\nu}H_{\mathrm{d}}\right)\rho_{\mathrm{d}}\right]},\\
    \beta_{\mu\nu}^{ij} & \equiv & 2\rho_{i}[\mathfrak{a}_{\mu}]_{ij}[\mathfrak{a}_{\nu}]_{ji},\nonumber
\end{eqnarray}
where $[\mathfrak{a}_{\mu}]_{ij}$ is the matrix element of $\mathfrak{a}_{\mu}$
in the representation whose basis are the eigenvectors of $H$. So
for such cases, $Q_{\mu\nu}^{(1)}$ is proportional to $t^{2}$, while
$Q_{\mu\nu}^{(2)}$ consists of circular functions. It is remarkable
that for pure states, $\alpha_{\mu\nu}$ is vanished, only the second
term of Eq.~(\ref{eq:Q_sep}) exists.

\section{XY model}
The one-dimension spin-1/2 XY model, with $N=2M+1$ spins, in the
presence of a transverse magnetic field characterized by $\lambda$,
is described by the Hamiltonian\begin{equation}
H_{\mathrm{XY}}=-\sum_{l=-M}^{M}\left[\frac{1+\gamma}{2}\sigma_{l}^{x}\sigma_{l+1}^{x}+\frac{1-\gamma}{2}\sigma_{l}^{y}\sigma_{l+1}^{y}+\lambda\sigma_{l}^{z}\right],\end{equation}
where $\sigma_{l}^{a}(a=x,y,z)$ are the Pauli matrix for the $l$-th
spin and $\gamma$ represents the anisotropy in the $x-y$ plane.
The periodic condition ($M+1=-M$) is assumed here. We are interested
in a new family of Hamiltonians obtained by applying a rotation of
$\phi$ around the $z$ direction to each spin\begin{equation}
H(\phi,\gamma,\lambda)=g^{\dagger}(\phi)H_{\mathrm{XY}}(\gamma,\lambda)g(\phi),\end{equation}
 where $g(\phi)=\prod_{l=-M}^{M}\exp\left(i\phi\sigma_{lz}/2\right)$.
This model was used to establish a relation between geometric phases
and criticality of spin chains \cite{Carollo2005,Zhu2006,Zhu2008}.

To obtain the analytic solution of this model, we apply three steps
of transformations as follows: (i) the Jordan-Wigner transformation
\cite{Lieb1961} $\sigma_{i}^{+}=\prod_{j<i}(1-2c_{j}^{\dagger}c_{j})c_{i}$,
$\sigma_{i}^{z}=1-2c_{i}^{\dagger}c_{i}$; (ii) Fourier transformation
$c_{l}=(1/\sqrt{N})\sum_{k}e^{i2\pi kl/N}d_{k}$ with $k=-M,\cdots, M$;
(iii) the pseudo Pauli operator transformation defined by \cite{Anderson1958}
\begin{eqnarray}
\sigma_{kx} & = & d_{k}^{\dagger}d_{-k}^{\dagger}+d_{-k}d_{k},\nonumber \\
\sigma_{ky} & = & -id_{k}^{\dagger}d_{-k}^{\dagger}+id_{-k}d_{k},\label{eq:psudospin}\\
\sigma_{kz} & = & d_{k}^{\dagger}d_{k}+d_{-k}^{\dagger}d_{-k}-1,\nonumber
\end{eqnarray}
which satisfy the commutative relations $\{\sigma_{ka},\sigma_{kb}\}=2\delta_{ab}$
and $[\sigma_{ka},\sigma_{kb}]=2i\varepsilon_{abc}\sigma_{kc}$, where
$\varepsilon_{abc}$ is the Levi-Civita symbol. For $k\neq0$, each
of them acts on the Fock space spanned by $\left\{ |\mathrm{vac}\rangle_{k,-k},|k,-k\rangle,|k\rangle,|-k\rangle\right\} $
with $|\mathrm{vac}\rangle$ the vacuum state for $d_{k}$ and $d_{-k}$
fermions. These pseudo Pauli operators become conventional Pauli matrix
in the subspace spanned by $\left\{ |0\rangle_{k,-k},|k,-k\rangle\right\} $
and vanish elsewhere.

After these three steps, we get
\begin{align}
    H & =(\lambda-1)\sigma_{0z}+\sum_{k=-M}^{M}H_{kz},\nonumber \\
    H_{kz} & =S_{k}^{\dagger}(\phi,\theta_{k})H_{k,\mathrm{d}}(\Lambda_{k})S_{k}(\phi,\theta_{k}),
\end{align}
where $H_{k,\mathrm{d}}=\Lambda_{k}\sigma_{kz}$ and $S_{k}(\phi,\theta_{k})=R_{kx}(\theta_{k})R_{kz}(\phi)$
with $R_{ka}(\alpha)=\exp(-i\alpha\sigma_{ka}/2)$ for $a=\{x,y,z\}$.
The intermediate parameters $\Lambda_{k}$ and $\theta_{k}$ are given
by
\begin{align}
\Lambda_{k} & =2\sqrt{\left(\lambda-\cos\left(2\pi k/N\right)\right)^{2}+\gamma^{2}\sin^{2}\left(2\pi k/N\right)},\nonumber \\
\theta_{k} & =-\frac{i}{2}\ln\frac{\lambda-\cos\left(2\pi k/N\right)+i\gamma\sin\left(2\pi k/N\right)}{\lambda-\cos\left(2\pi k/N\right)-i\gamma\sin\left(2\pi k/N\right)}.
\end{align}

We consider the OQGT of the time evolution operators $U=\exp(-itH)$
and chose the ground state $\rho=|G\rangle\langle G|$ as the reference
state. They are of the same factorization structure
\begin{align}
U & =\bigotimes_{k=0}^{M}S_{k}^{\dagger}(\phi,\theta_{k})\exp(-itH_{k,d})S_{k}(\phi,\theta_{k}),\nonumber \\
\rho & =\bigotimes_{k=0}^{M}S_{k}^{\dagger}(\theta_{k})|\downarrow\rangle_{k}\langle\downarrow|S_{k}(\theta_{k}),\end{align}
where $|\downarrow\rangle_{k}$ is the eigenstate of $\sigma_{kz}$
with eigenvalue $-1$. Due to this factorization structure, the OQGT
is additive. Because $[\rho,U]=0$ and $\rho$ is pure state, $Q_{\mu\nu}$
is splitted and $Q_{\mu\nu}^{(1)}$ of Eq.~(\ref{eq:OQGT_time_evolution})
vanishes. To obtain the $Q_{\mu\nu}^{(2)}$, we first get the 1-form
matrix $\mathfrak{a}\equiv iS^{\dagger}dS$, which is given by $\mathfrak{a}=\sum_{k=0}^{M}\mathfrak{a}_{k}$
with
\begin{eqnarray}
\mathfrak{a}_{k} & = & iS_{k}^{\dagger}(\phi,\theta_{k})d\left[S_{k}(\phi,\theta_{k})\right]\nonumber \\
 & = & \frac{d\theta_{k}}{2}R_{kz}^{\dagger}(\phi)\sigma_{kx}R_{kz}(\phi)+\frac{d\phi}{2}\sigma_{kz}.\label{eq:a_k}\end{eqnarray}
Substituting Eq.~(\ref{eq:a_k}) into Eq.~(\ref{eq:OQGT_time_evolution}),
we obtain the components of the OQGT as follows
\begin{eqnarray*}
    Q_{\lambda\lambda} & = & \sum_{k=1}^{M}\frac{4}{\Lambda_{k}^{2}}\sin^{2}\left(\Lambda_{k}t\right)\sin^{2}\theta_{k},\\
    Q_{\gamma\gamma} & = & \sum_{k=1}^{M}\frac{4}{\Lambda_{k}^{2}}\sin^{2}\left(\Lambda_{k}t\right)\cos^{2}\theta_{k}\sin^{2}\left(\frac{2\pi k}{N}\right),\\
    Q_{\phi\phi} & = & \sum_{k=1}^{M}\sin^{2}\left(\Lambda_{k}t\right)\left[\sin\left(2\theta_{k}\right)\right]^{2},\\
    Q_{\lambda\gamma} & = & \sum_{k=1}^{M}-\frac{4}{\Lambda_{k}^{2}}\sin^{2}\left(\Lambda_{k}t\right)\sin\theta_{k}\cos\theta_{k}\sin\left(\frac{2\pi k}{N}\right),\\
    Q_{\lambda\phi} & = & \sum_{k=1}^{M}-i\frac{2}{\Lambda_{k}}\sin^{2}\left(\Lambda_{k}t\right)\sin\theta_{k}\sin\left(2\theta_{k}\right),\\
    Q_{\gamma\phi} & = & \sum_{k=1}^{M}i\frac{2}{\Lambda_{k}}\sin^{2}\left(\Lambda_{k}t\right)\cos\theta_{k}\sin\left(2\theta_{k}\right)\sin\left(\frac{2\pi k}{N}\right).
\end{eqnarray*}
With the help of the metric tensor $g_{\mu\nu}\equiv\mathrm{Re}\, Q_{\mu\nu},$
we can obtain the Loschmidt echo $L=|\langle G(x)|U^{\dagger}(x+\delta x)U(x)|G(x)\rangle|^{2}$
for arbitrary multi-parameter perturbations $\delta x=(\cdots,\delta x^{\mu},\cdots)$.
This can be seen from Eq.~(\ref{eq:distance}), then we have $L=1-g_{\mu\nu}\delta x^{\mu}\delta x^{\nu}/4$.
So the metric tensor directly reflects the sensitivity of the time
evolution to the perturbations of multi parameters. A relation between
the Loschmidt echo and quantum criticality was established in Ref.~\cite{Quan2006}.

\begin{figure*}
    \begin{center}
        \includegraphics[scale=0.7]{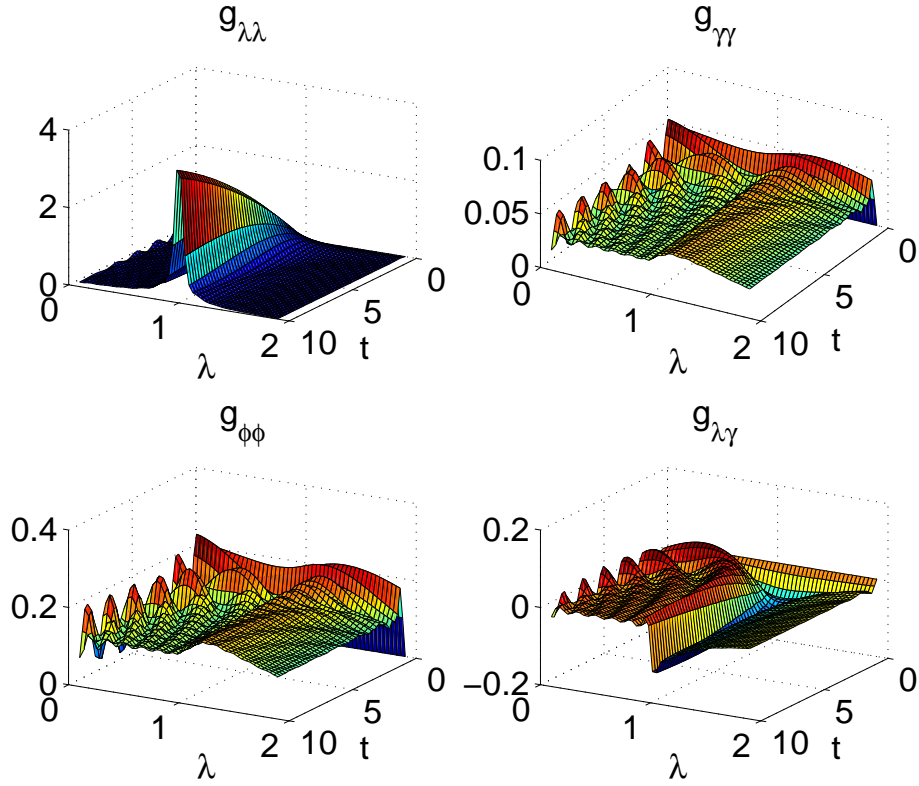}
    \caption{Three dimensional diagram of the components of the rescaled metric
    tensor $g_{\mu\nu}=\mathrm{Re\, Q_{\mu\nu}/N}$ as functions of $\lambda$
    and $t$ for the system at $\gamma=1$ with $N=1001$ the number of
    spins. Here $\mu$,$\nu$ are chosen from the parameter set $\{\lambda,\gamma,\phi\}$. }
    \label{fig:metric_tensor}
    \end{center}
\end{figure*}

The critical points of the XY model are given by the lines $\lambda=\pm1$
and by the segment $|\lambda|<1,\,\gamma=0$ \cite{Sachdev1999}.
We consider the region of the Ising-type phase transition at $\lambda=1$.
The time evolutions of the components of the metric tensor $g_{\mu\nu}=\mathrm{Re}\, Q_{\mu\nu}$
for different values of $\lambda$ are shown in Fig.~(\ref{fig:metric_tensor})
, at $\gamma=1$ for simplicity. The behaviors of the $\lambda$-related
components $g_{\lambda\lambda}$ and $g_{\lambda\gamma}$ are obviously
singular at the critical point $\lambda=1$. $g_{\lambda\lambda}$
abruptly increases with time near the critical point, which implies
the time evolution is highly sensitive in this critical point. Near
the critical point, $g_{\lambda\gamma}$ also has a qualitative distinction,
it sharply decreases to negative values with time when $\lambda$
slightly greater than 1.

\begin{figure*}
    \begin{center}
        \includegraphics[scale=0.8]{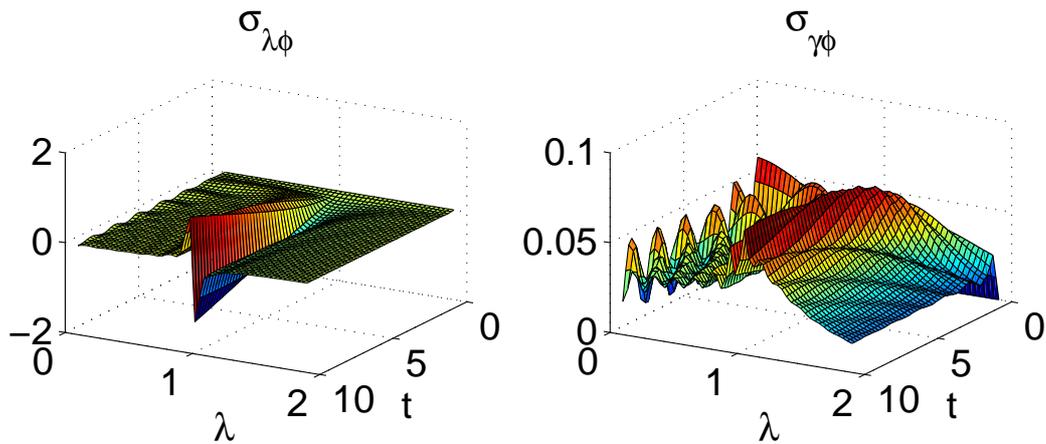}
        \caption{Three dimensional diagram of the components of the rescaled curvature
        tensor $\sigma_{\mu\nu}=\mathrm{Im\, Q_{\mu\nu}/N}$ as functions
        of $\lambda$ and $t$ for the system at $\gamma=1$ with $N=1001$  the number of
    spins. Here $\mu$,$\nu$ are chosen from the parameter set $\{\lambda,\gamma,\phi\}$.}
        \label{fig:curvature_tensor}
    \end{center}
\end{figure*}

In the three-dimensional manifold of the parameter $(\lambda,\gamma,\phi)$,
the curvature 2-form $\sigma$ is equivalent to a effective vector
field $\mathbf{B}=(B_{\lambda},B_{\gamma},B_{\phi})$ through the
relation
\begin{equation}
    \sigma=B_{\lambda}d\lambda^{\gamma}\wedge d\lambda^{\phi}+B_{\gamma}d\lambda^{\phi}\wedge d\lambda^{\lambda}+B_{\phi}d\lambda^{\lambda}\wedge d\lambda^{\gamma}.
\end{equation}
In this rotated XY model, we obtain $\mathbf{B}=(2\mathrm{Im}Q_{\gamma\phi},-2\mathrm{Im}Q_{\lambda\phi},0)$.
For the case of $\gamma=1$, the time evolution of $B_{\gamma}$ ($-2\sigma_{\lambda\phi}$)
is very different between the two sides of the critical point, as shown in Fig.~(\ref{fig:curvature_tensor}). When
$\lambda$ is slightly less than 1, $B_{\gamma}$ sharply increases with
time, meanwhile when $\lambda$ is slightly greater than 1, $B_{\gamma}$ sharply
decreases from zero to negative values. So we conclude that the amplitude
of $B_{\gamma}$ sharply increases near the critical but the direction
of $B_{\gamma}$ is opposite in the different phases.

\section{Conclusion}
To summarize, we have introduced the operator-quantum-geometric tensor
on the manifold of the unitary operators with a given reference state,
and prove the additivity for factorizable models. We show the splitting
of two kinds contributions to the OQGT for the cases of stationary
reference states, and these two contributions have different type
of time dependence. This OQGT approach can be applied to investigate
the sensitivity of unitary operation against perturbations. The Loschmidt
echo for the unperturbed and perturbed time evolutions for arbitrary
kind of multi-parameter perturbations can be easily got as long as
the OQGT is obtained. We used this approach to investigate sensitivity
of the time evolutions of the rotated XY models and show
the sharp changes of the OQGT near the quantum critical points. We
believe that this OQGT approach is useful to some interesting questions,
like the quantum criticality \cite{Wang2009,Sun2009}, decoherence
\cite{Lu2008} and quantum chaos \cite{Giorda2010,TobiasJacobson2010}, etc.
After the acceptance of this paper, we notice the work~\cite{Rezakhani2010} where both the QGT and Operator fidelity  are shown to be relevant to the adiabatic error in the adiabatic and holonomic quantum computing.

\acknowledgments
The authors thank P.~Zanardi, S.-J. Gu, Z.~D.~Wang, and C.~P.~Sun for
the helpful discussions and communication.
This work is supported by NSFC with grant No.\ 10874151, 10935010,
NFRPC with grant No.\ 2006CB921205,  and the Fundamental Research
Funds for the Central Universities.

\end{document}